\documentclass[sigconf, review=false]{acmart}

\usepackage{booktabs}
\usepackage{graphicx}
\usepackage{listings}
\usepackage{xcolor}
\newcommand{\bad}{\textcolor{red}{\textbf{\texttimes}}}
\newcommand{\good}{\textcolor{green!60!black}{\textbf{\checkmark}}}
\newcommand{\yellow}{\textcolor{yellow!80!black}{\checkmark}}
\usepackage{caption}
\usepackage{float}
\usepackage{enumitem}
\usepackage{needspace}

\makeatletter
\let\ACM@line@margin\relax
\let\ACM@lineno\relax
\makeatother

\begin{document}
\settopmatter{printacmref=false}

% Title and author information
\title{Mobile App Rewrites via Dual Boot}
\subtitle{Dynamic Library Partitioning, In-Binary Experimentation, and Graceful Deprecation}

\author{Gaurav Nijhara}
\affiliation{
  \institution{Meta Platforms, Inc.}
    \country{USA}
}

\author{Jasmit Kaur Saluja}
\affiliation{
  \institution{Meta Platforms, Inc.}
   \country{USA}
}

\author{Pritesh Shah}
\affiliation{
  \institution{Meta Platforms, Inc.}
    \country{USA}
}

\author{Dustin Shahidehpour}
\affiliation{
  \institution{Meta Platforms, Inc.}
    \country{USA}
}

\author{Hari Panjwani}
\affiliation{
  \institution{Meta Platforms, Inc.}
    \country{USA}
}

% Keywords
\keywords{mobile architecture, dynamic linking, code sharing, strangler fig pattern, A/B testing, application migration, dual boot}

\begin{abstract}
We term the mechanism \textit{dual boot}: the ability to host two application variants within a single binary with a boot-time variant selection. This mechanism, combined with build-generated symbol resolution maps and linker retain lists, also enables two additional capabilities not inherent to the mechanism alone: (1) in-binary A/B experimentation between full application variants, and (2) graceful deprecation of the legacy variant without user disruption: preserving app store listing, branding, and install base. We frame the complete lifecycle as the first application of the Strangler Fig pattern to native mobile platforms. We validate the architecture across three evaluation scenarios, with improvements in developer productivity, experimentation capability and flexibility. The system contributed to a complete iOS application rewrite, from initial embedding through multi-segment experimentation to full deprecation of the legacy variant, without losing users throughout the 10-month evaluation.
\end{abstract}

\maketitle

\section{Introduction}

Large-scale mobile applications periodically require fundamental transitions: a full rewrite of the underlying architecture, a new product experience built on shared infrastructure, or a migration from one technology stack to another. These transitions present complexities. A new variant must be validated with real users before the organization can commit to deprecating the original, yet the original must remain fully operational throughout the validation period.

Existing approaches force a choice between isolation and reuse. Publishing a separate application provides clean separation but sacrifices the existing install base, ratings, and distribution channel. Embedding conditional logic within the existing codebase preserves distribution but entangles the two variants at the dependency level, which can make eventual separation difficult and error-prone.

This architecture transforms a mobile application rewrite from a major undertaking into an iterative, measurable, reversible process. By embedding the new variant as a dynamic framework within the existing binary, organizations can validate a complete rewrite with real users, iterate based on metrics, and retain the ability to abandon or course-correct at any point with user impact minimized. The technical basis enabling (dylib partitioning, build-generated symbol resolution maps, and a custom runtime loader) is what makes controlled rollout of a full application rewrite possible within a single distributable binary.

We propose \textit{Dual Boot with Code Sharing}: an architecture achieving build-time isolation between application variants while enabling runtime code sharing through dynamic libraries. Our contributions:

\begin{itemize}
    \item A dylib-partitioned architecture with true build-time isolation while sharing common frameworks (\S3).
    \item A generalized build pipeline generating entry-point-to-dylib maps, solving symbol stripping for any runtime extension mechanism, with formal algorithms enabling reproduction (\S4).
    \item An in-binary experimentation framework enabling variant-level A/B testing and a graceful deprecation path preserving app identity (\S5).
    \item A multi-pipeline CI/CD architecture with centrally generated shared resources (\S6).
    \item The first formalization of the Strangler Fig pattern for native mobile platforms (\S2).
    \item Validation across three evaluation scenarios (\S7).
\end{itemize}

\section{Background and Related Work}

\subsection{Dynamic Linking and Runtime Extension}

Dynamic linking enables applications to load code at runtime rather than requiring all code to be statically linked at compile time. On iOS, dynamic frameworks are loaded via \texttt{dlopen()} and their symbols resolved via \texttt{dlsym()} \cite{ref1,ref3}. On Android, the ClassLoader and Dynamic Feature Modules \cite{ref4, ref9} provide equivalent functionality \cite{ref4}. These mechanisms allow mobile applications to organize code into independently compiled modules that are assembled at runtime, a property our architecture leverages to host multiple application variants within a single binary.

The implementation presented in this paper targets iOS and uses dynamic frameworks as its loading mechanism. The underlying principles of runtime code loading, symbol resolution, and build-time partitioning are not specific to any one platform. Android provides analogous capabilities through ClassLoader and Dynamic Feature Modules, though the system described here was evaluated exclusively on iOS.

\subsection{Prior Approaches}

\begin{table}[h]
\centering
\caption{Comparison of prior approaches}
\begin{tabular}{@{}p{1.5cm}p{2.6cm}p{3.4cm}@{}}
\toprule
\textbf{Approach} & \textbf{Mechanism} & \textbf{Limitation} \\
\midrule
UI Branching & Conditional logic throughout codebase & No build isolation; branch pollution; entanglement \\
\addlinespace
App Extensions (iOS) & Separate processes with IPC & Designed for widgets and share sheets, not full app variants; cannot share in-memory state or UI framework across process boundary \\
\addlinespace
Android Dynamic Features & On-demand module delivery via Play Store & Designed for optional feature delivery, not for hosting a complete alternative app experience; no variant-level experimentation support \\
\addlinespace
OSGi (Java) & Bundle lifecycle and service registry & JVM-only; runtime \cite{ref5} \\
\addlinespace
Separate App Listing & Independent binary, shared backend only & User migration required; lost install base and ratings \\
\bottomrule
\end{tabular}
\end{table}

\subsection{The Strangler Fig Pattern and Mobile Constraints}

The Strangler Fig pattern (Fowler, 2004) \cite{ref7} describes an incremental migration strategy: wrap new functionality around the existing system, progressively redirect traffic, and eventually decommission the original. This is well-established for server-side monolith-to-microservices migrations \cite{ref7, ref15}.

However, it has not been formally applied to \textit{native mobile applications}, which face constraints absent from server-side systems:

\begin{itemize}
    \item \textbf{Single-binary distribution:} Mobile apps ship as one artifact through app stores: you cannot independently release new endpoints alongside old ones.
    \item \textbf{App identity coupling:} Bundle identifiers, store listings, and ratings are tied to the binary: a new app means starting from zero.
    \item \textbf{Linker optimization:} Server-side code is not subject to dead-code elimination. On mobile, both iOS and Android aggressively optimize binaries: Apple's linker strips unreferenced symbols by default, and Android's R8 optimizer removes unreferenced classes: making runtime-discovered features vulnerable to removal in any multi-module app.
    \item \textbf{Client-side state:} Mobile apps carry local data that must persist through transitions.
\end{itemize}

Our architecture is the Strangler Fig pattern adapted for these constraints: the legacy app (App A) is the strangled system; the new app (App B) grows around it within the same binary; the Shared Framework provides the stable shared dependencies both variants build upon; experiment configuration at launch progressively redirects users between variants; and deprecation removes the legacy dylib: all while preserving app identity.

\subsection{Industrial Practice}

\begin{table}[h]
\centering
\caption{Industrial practice comparison}
\label{tab:industrial-practice}
\begin{tabular}{@{}p{1.5cm}p{2.5cm}p{3.5cm}@{}}
\toprule
\textbf{Company} & \textbf{Strategy} & \textbf{Limitation vs. Dual Boot} \\
\midrule
Uber: Rider App (2016) \cite{ref8} & Ground-up rewrite with Riblets architecture \cite{ref8}; big-bang cutover & No gradual rollout; no A/B testing between old and new architectures; no instant rollback to the previous version \cite{ref8} \\
\addlinespace
Airbnb: MvRx Migration (2018) & Incremental UI rewrite via feature flags \cite{ref10} & Feature flags accumulate conditional logic proportional to the number of migrated surfaces; removal of deprecated paths requires auditing shared code for residual references \\
\addlinespace
Microsoft: Teams Mobile (2022) & Ground-up rewrite; shipped as update \cite{ref11} & Users experienced abrupt feature gaps; no side-by-side comparison period \\
\bottomrule
\end{tabular}
\end{table}

None of these approaches provide the combination of: (a) build-time isolation, (b) runtime sharing without IPC, (c) in-binary experimentation, and (d) seamless deprecation preserving app identity. Dual boot achieves all four.

Specifically, our contributions differ from prior industrial practice as follows. Unlike Uber's big-bang rewrite \cite{ref8}, our architecture supports gradual rollout with instant rollback and variant-level A/B testing throughout the transition. Unlike Airbnb's incremental migration \cite{ref10}, our approach avoids accumulating conditional branching in shared code, instead providing structural separation at the compilation unit level. Unlike Microsoft Teams' rewrite \cite{ref11}, users experienced no abrupt feature gaps because both variants ran simultaneously with progressive migration. Prior work on large-scale mobile refactoring~\cite{ref13} has examined modularization strategies but not the full embed-experiment-deprecate lifecycle."

\section{System Architecture}

\subsection{Overview}

Consider two mobile applications: App A (the existing product) and App B (a new variant or rewrite). Each is developed as an independent codebase with its own libraries, build targets, and CI pipeline. Neither has compile-time visibility into the other; a developer working on App B never encounters App A's code.

The dual boot architecture merges these two independent applications into a single distributable binary. App B is compiled as a dynamic framework and bundled alongside App A during a combined build step. At launch, the binary reads a cached experiment assignment and loads either App A or App B's framework. The decision happens once, at the process entry point, before any application-level code runs. From the user's perspective, they have one app that behaves as whichever variant they are assigned to.

Because both apps ship within one bundle, they share a single app store listing, a single bundle identifier, and a single set of centrally-generated resources (localization tables, experiment parameters, asset catalogs). The architecture enables variant-level A/B testing where the experiment system assigns users to App A or App B, while maintaining complete code-level independence between the two.

The architecture has four phases:

\textbf{Build time} (\S4): Two application variants are compiled into separate dynamic libraries sharing a common framework. A build pipeline extracts all runtime-resolvable runtime-resolved symbols into a unified manifest and generates a symbol-to-dylib resource map. These artifacts serve both app variants within the single bundle. Similarly, localization tables, experiment parameters, and asset catalogs are generated centrally by merging across all dylib targets, since a single app bundle requires one unified set of each. The manifest also produces a retain list injected into the linker to prevent symbol stripping.

\textbf{Distribution:} The resulting binary ships as a single artifact through the app store. Same bundle ID, same listing, same brand as before. Users see a normal app update.

\textbf{Runtime} (\S4.3, \S5): At launch, an experiment configuration determines which variant to load. The selected dylib is loaded via \texttt{dlopen}. A DylibLoader component reads the bundled resource map and resolves all cross-dylib feature calls: enabling each variant to invoke shared infrastructure and (where permitted) cross-reference runtime-resolved symbols in other loaded dylibs. Experiment metrics (retention, engagement, crash rates) are collected per-variant for comparison.

\textbf{Deprecation} (\S5.3): Once metrics validate the new variant, the legacy dylib is removed from the build configuration. The next release ships a single-variant binary through the same app store listing. No user migration, no new bundle ID, no lost ratings. The transition is invisible to end users.

The following subsections detail each architectural component, following established architectural documentation practices~\cite{ref12}"

\begin{figure}[h]
  \centering
  \includegraphics[width=\columnwidth]{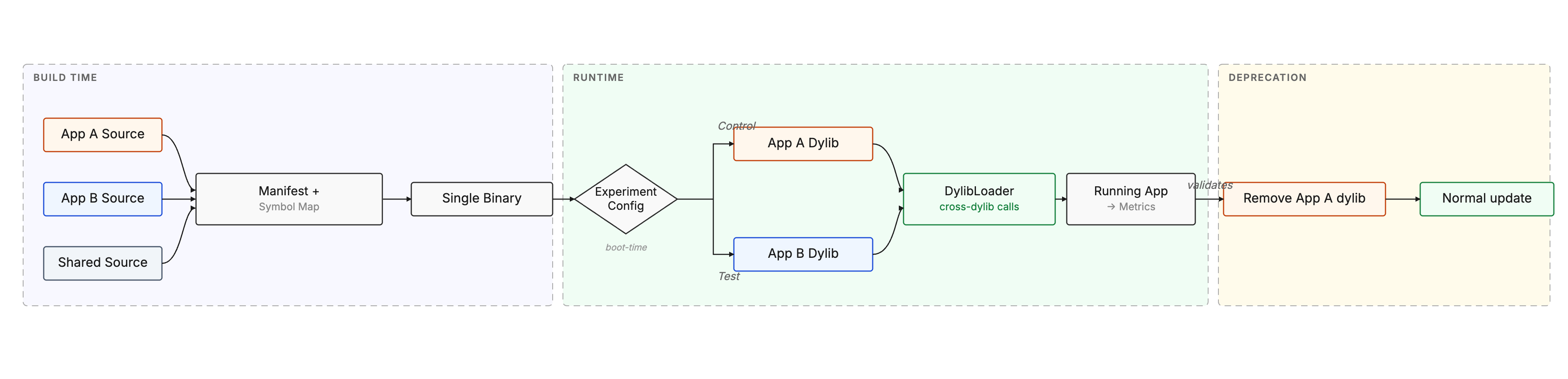}
  \caption{End-to-end lifecycle.}
  \label{fig:end-to-end-lifecycle}
\end{figure}

\subsection{Dylib Partitioning Model}

The system partitions the shared codebase into dynamic libraries with distinct roles and lifecycles:

\textbf{Startup Set:} Frameworks loaded unconditionally at launch. Contains infrastructure both variants require (networking, encryption, persistence, telemetry). Changes here affect both apps.

\textbf{Shared Framework:} Common application-level code both variants depend on. Defines a stable API: this is the boundary that enables isolation. Examples: UI component library, media pipeline, synchronization engine. Loaded via \texttt{dlopen} during early initialization.

\textbf{App-Specific Dylibs:} Each variant's unique code resides in its own dylib, loaded conditionally based on runtime context. These have \textit{no compile-time visibility} into each other: the build system enforces this structurally, not by convention \cite{ref14}.

\begin{figure}[h]
  \centering
  \includegraphics[width=\columnwidth]{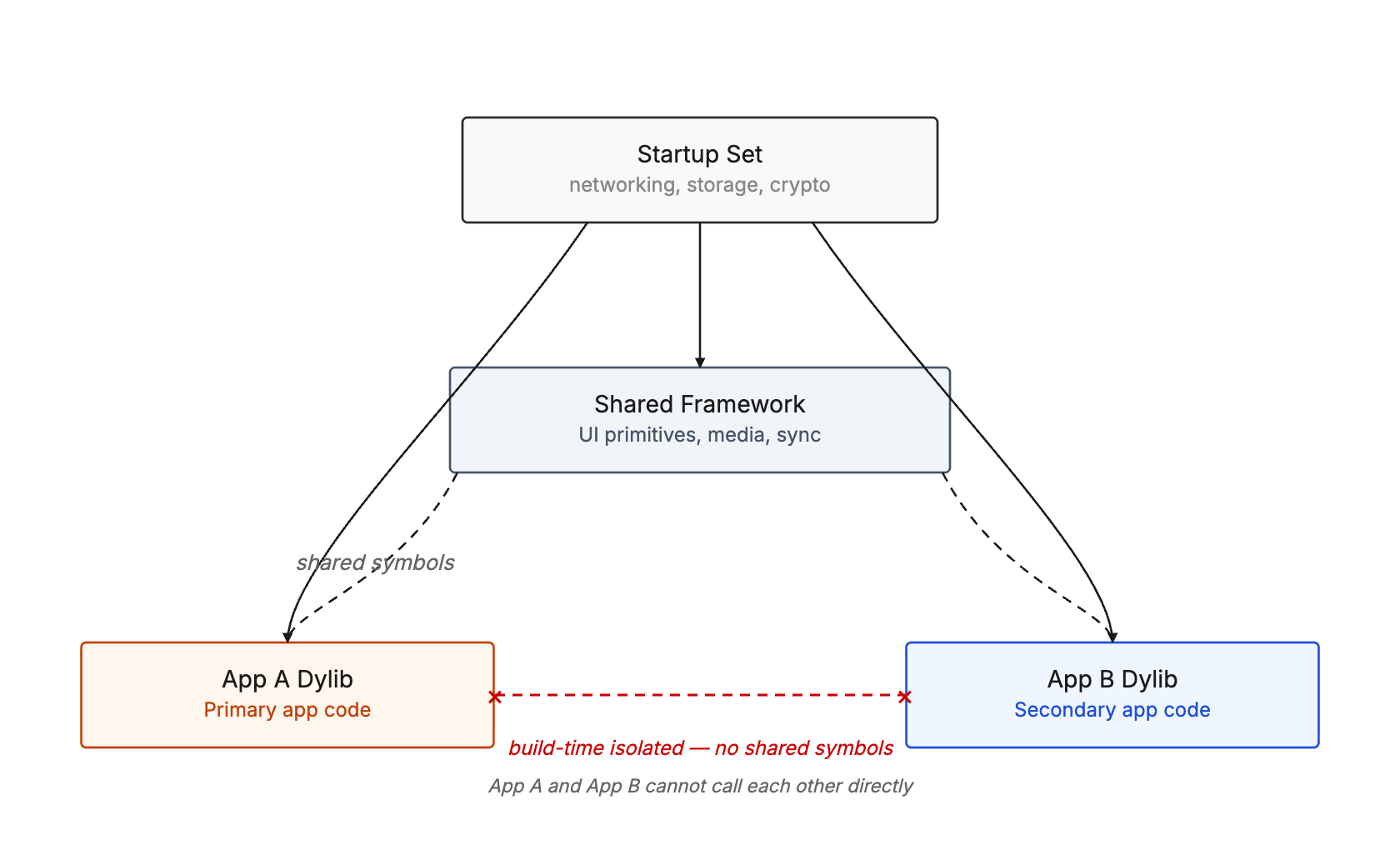}
  \caption{Dylib partitioning model.}
  \label{fig:dylib-partitioning}
\end{figure}

\subsection{Runtime Initialization}

\begin{figure}[h]
  \centering
  \includegraphics[width=\columnwidth]{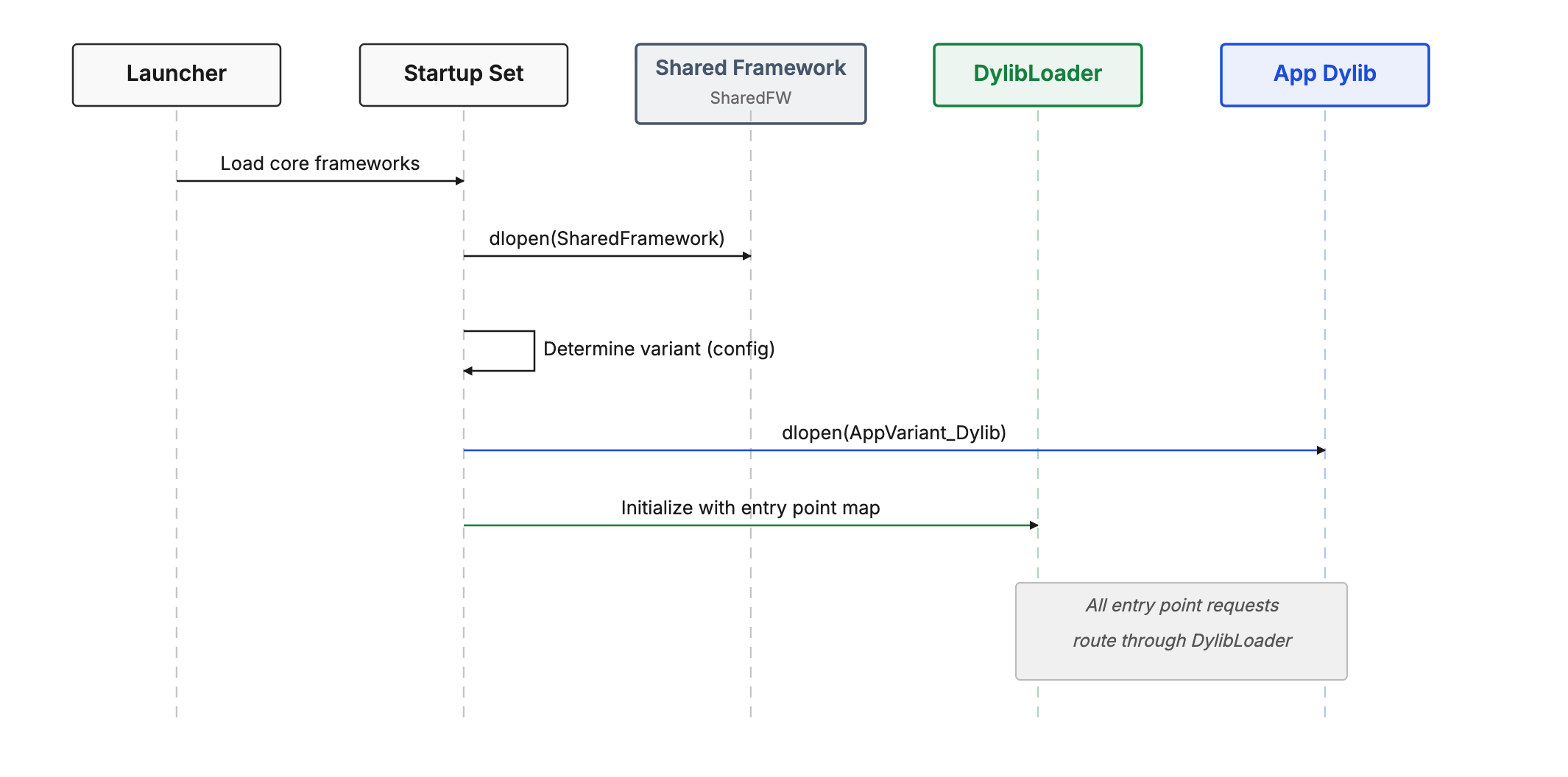}
  \caption{Runtime initialization sequence.}
  \label{fig:runtime-initialization}
\end{figure}

\subsection{Shared Framework}

The Shared Framework is a dynamic library containing infrastructure and UI components that both variants depend on: networking, storage, encryption, theming, and layout primitives. Because this code is compiled once and loaded by both App A and App B at runtime, it eliminates duplication that would otherwise nearly double binary size. Both variants invoke the Shared Framework, but neither variant may invoke the other. The interface is versioned and validated independently per variant at build time. If App B references a Shared Framework symbol whose signature has changed, the App B build fails without affecting App A.

\section{Runtime Symbol Resolution Across Dylib Boundaries}

In a single-application build, all code resides in one compilation unit and the compiler resolves every function call by address at link time. A multi-dylib layout removes this guarantee. When the active variant invokes a symbol that resides in a separately compiled dylib, that call cannot be resolved statically because the target was not visible to the linker at build time. It must instead be resolved by name at runtime. For example, suppose App A contains a screen that displays a media gallery. The rendering logic resides in App A's dylib, but the image decoding function it calls lives in the Shared Framework, which was compiled as a separate dynamic library. At link time, App A's binary has no static reference to that decoder symbol. When the gallery screen executes, the call must be resolved at runtime by loading the Shared Framework's dylib and looking up the symbol by name.

Any application that partitions code across dynamically loaded libraries faces this resolution requirement: cross-dylib symbol calls must be resolved by name at runtime. Our codebase additionally employed named function pointers as a modularity construct, meaning that feature dispatch relied on runtime resolution even within a single variant. This second pattern is specific to codebases that already use function-pointer-based dispatch and may not apply to all adopters of the architecture.

This section examines why name-based resolution introduces a correctness challenge in optimized builds and presents a build pipeline that addresses it. The problem generalizes to any multi-dylib mobile application regardless of its internal dispatch architecture.

\subsection{The Symbol Stripping Problem}

Modern mobile toolchains apply aggressive dead-code elimination (DCE): the linker strips any symbol not transitively referenced from known runtime-resolved symbols at compile time. This creates a problem for modular architectures where features are discovered at runtime.

Consider: App B registers a feature initialization function as a named function pointer. At runtime, a caller within App B's own dylib resolves this pointer through the DylibLoader via \texttt{dlsym(handle, "FeatureInitFunc")}. Because no compile-time call graph edge references \texttt{"FeatureInitFunc"}, the linker treats it as dead code and eliminates it. The app crashes at runtime when the resolution returns a null pointer. This affects any runtime extension mechanism. The linker optimizes binary size by removing symbols that have no static references. When an architecture resolves calls by name at runtime, those symbols appear unreferenced at link time and are removed. The system therefore requires a mechanism to declare symbols that must be retained despite the absence of compile-time references.

\subsection{Solution: Build-Generated Symbol Resolution Maps}

We solve this with a build pipeline that generates two artifacts: (1) a \textit{symbol-to-dylib resource map} bundled into the app for runtime resolution, and (2) a \textit{retain list} (also called an ``export list'' or ``keep list'') fed to the linker to prevent stripping.

\begin{figure}[h]
  \centering
  \includegraphics[width=\columnwidth]{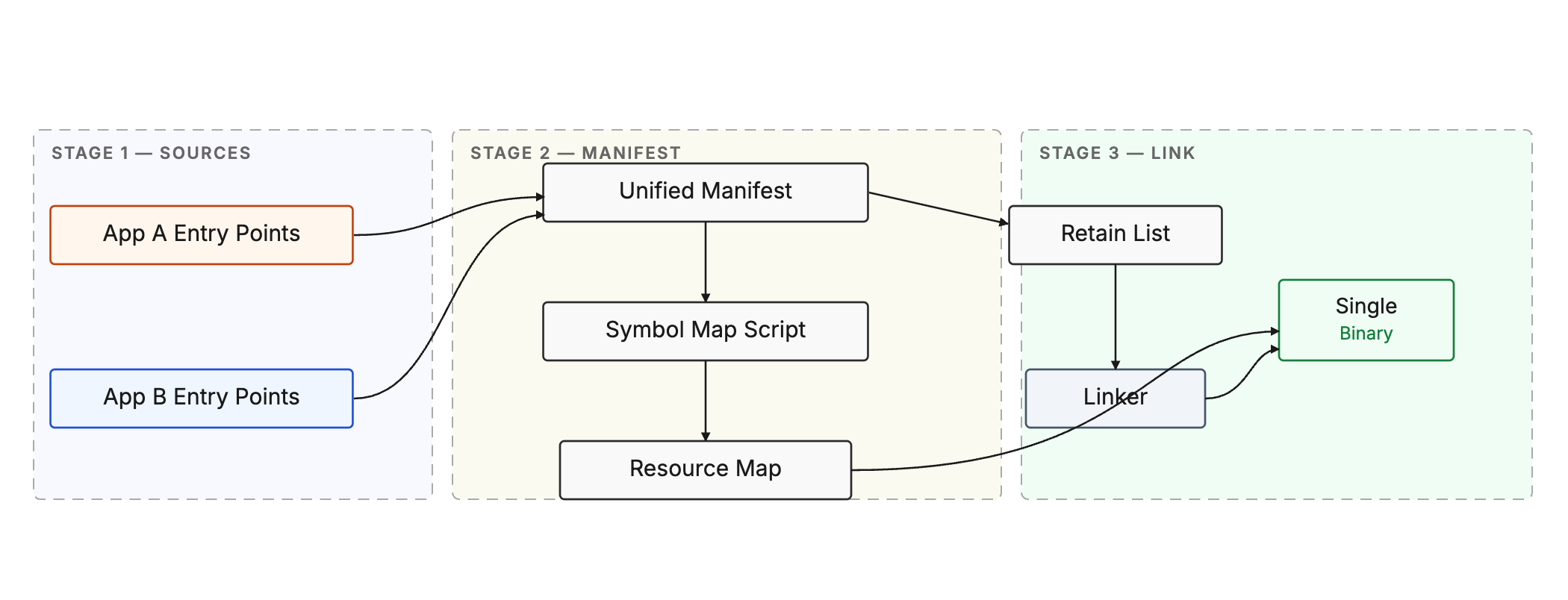}
  \caption{Build pipeline for symbol map generation.}
  \label{fig:build-pipeline}
\end{figure}

\subsection{The DylibLoader}

At runtime, a component we call the \textit{DylibLoader} acts as a \textit{service locator \cite{ref2} that knows which dynamic library contains each runtime-resolved symbol}. When code requests a feature by name, the DylibLoader:

\begin{enumerate}
    \item Checks its cache (O(1) for repeated calls).
    \item Looks up the runtime-resolved symbol name in the bundled resource map $\rightarrow$ gets the dylib identifier.
    \item Ensures that dylib is loaded (calls \texttt{dlopen} if it hasn't been loaded yet).
    \item Resolves the symbol from the correct dylib via \texttt{dlsym}.
    \item Caches the resolved pointer for all future calls.
\end{enumerate}

\begin{figure}[h]
  \centering
  \includegraphics[width=\columnwidth]{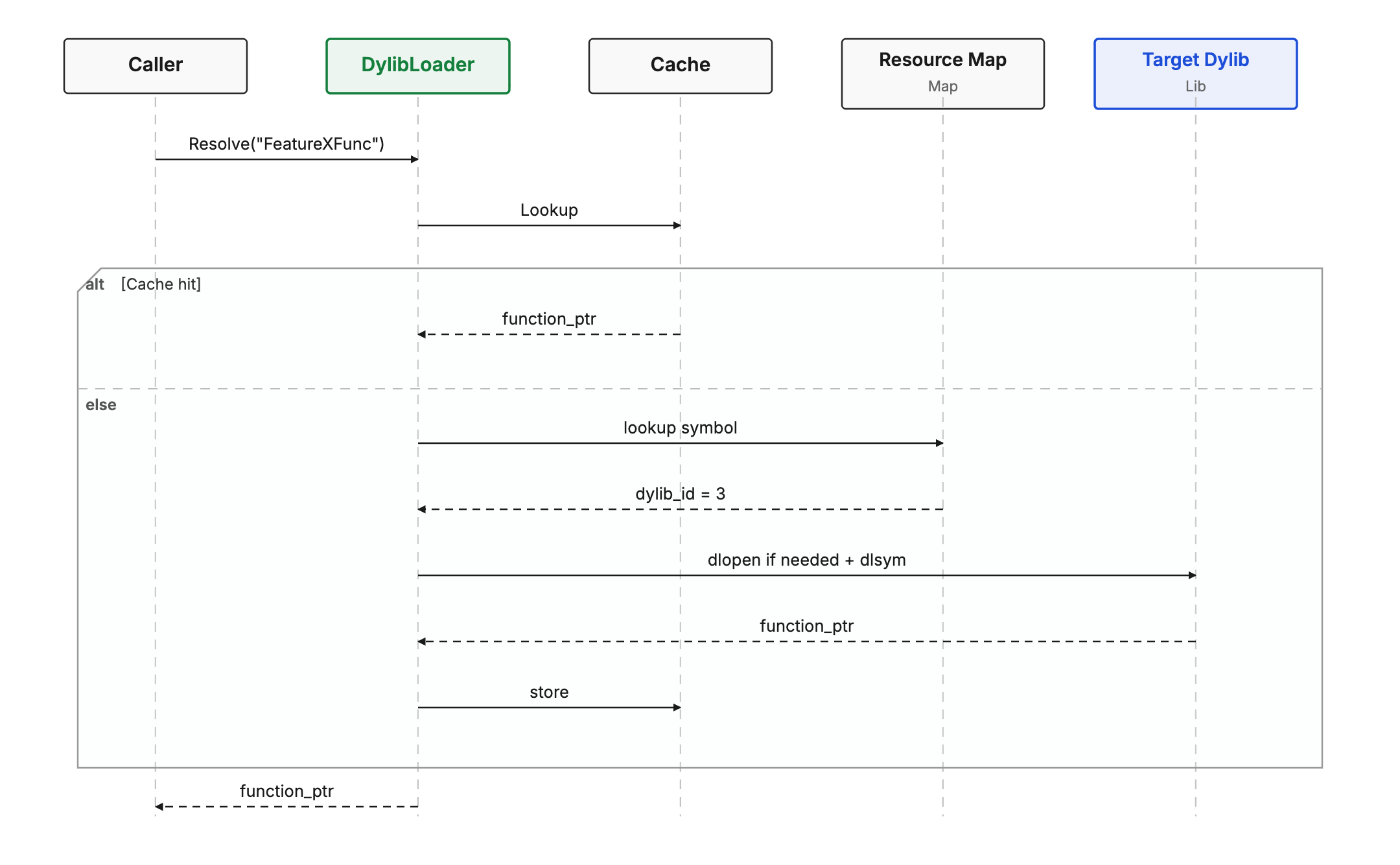}
  \caption{DylibLoader resolution with caching.}
  \label{fig:dylibloader-resolution}
\end{figure}

\subsection{Cross-Dylib Validation}

A runtime-resolved symbol's interface may reside in a different dylib than its implementation. For example, a navigation interface defined in the Shared Framework may have an implementation in App B's dylib. The build system validates these pairs:

\textbf{Cross-dylib safe:} The DylibLoader handles resolution at runtime. Build validates the implementation's dylib will be available.

\textbf{Co-location required:} Interface and implementation must share a dylib. Violations are build errors: catching at compile time what would otherwise be a runtime crash.

\section{Experimentation and Graceful Deprecation}

\subsection{In-Binary A/B Testing}

An advantage of the dual boot architecture is its native support for \emph{in-binary experimentation}. Because both application variants exist within the same distributable binary, the system can assign users to control (App A) or test (App B) groups without requiring separate app installations or app store listings. The experiment assignment occurs at process launch (in \texttt{main()}), before application-level initialization.

On a user's first launch, the binary loads App A unconditionally. During that initial session, the experiment system contacts the server, determines the user's group assignment, and writes it to local storage. Every subsequent launch reads this persisted assignment in \texttt{main()}, before any framework or application code initializes, and loads the corresponding variant dylib. Because the decision reduces to a single local disk read, it completes in negligible time and does not depend on network connectivity or framework initialization order.

\textbf{Experiment segments} enabled by this architecture:

\begin{itemize}
    \item \textbf{Forced test group:} New users and priority segments are assigned directly to App B with no opt-out, enabling clean metrics without legacy bias.
    
    \item \textbf{Opt-in/Opt-out group:} Existing users receive an in-app prompt to try App B, with the ability to switch back to App A at any time, enabling preference measurement.
    
    \item \textbf{Holdout group:} A reserved percentage remains on App A permanently, enabling long-term causal measurement of App B's impact.
\end{itemize}

\begin{figure}[htbp]
    \centering
    \includegraphics[width=\columnwidth]{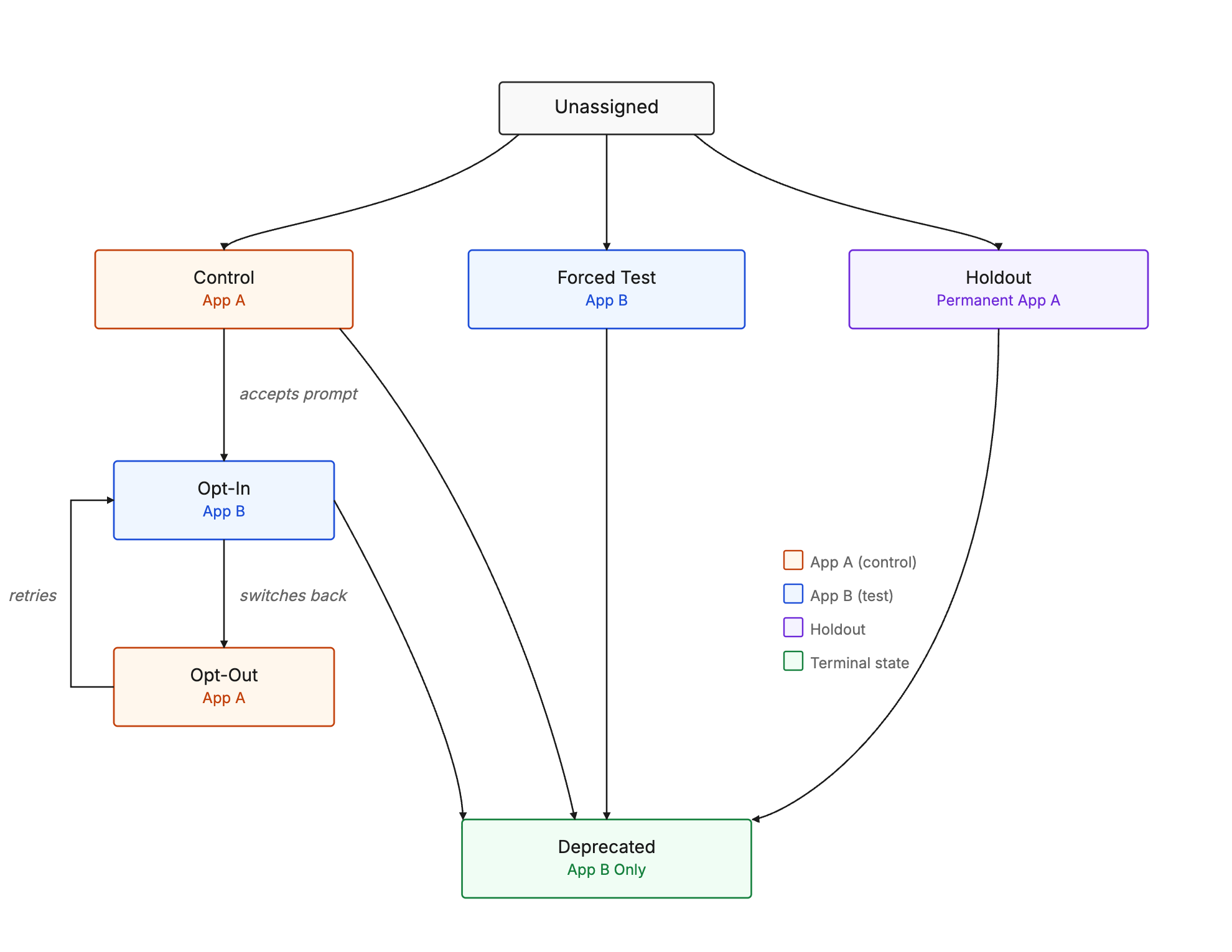}
    \caption{Experiment state machine. Users progress from unassigned through control/test groups toward deprecation.}
    \label{fig:experiment-state-machine}
\end{figure}

\subsection{Addressing Exposure Imbalance}

A subtle challenge arises because experiment assignment happens at launch but the configuration system initializes \emph{after} the boot decision. On their first launch, test-group users briefly experience the control variant before their assignment is cached. To maintain experimental validity:

\begin{itemize}
    \item The first-launch session is excluded from analysis for both groups.
    \item Exposure is logged symmetrically on the second launch for both control and test groups.
\end{itemize}

This ensures balanced exposure counts despite the one-session assignment delay.

\subsection{Phased Rollout and Evaluation Strategy}

The experiment did not follow a standard equal-split A/B test design. Instead, the architecture enabled a phased migration with multiple concurrent user segments, each gated independently.

\textbf{Wave-based expansion.} Users were migrated to the new variant in scheduled waves, expanding incrementally (1\% $\rightarrow$ 10\% $\rightarrow$ 50\% $\rightarrow$ 100\%). Pauses between waves allowed metric evaluation and regression fixes before proceeding.

\textbf{Concurrent segment types.} At any point, the user population comprised: forced test (new users assigned directly to App B), opt-in (users who voluntarily switched), opt-out (users who reverted to App A), not-yet-migrated (remaining on App A as de facto control), and a permanent holdout (never exposed to App B).

\textbf{Evaluation with unequal groups.} Because the test group grew incrementally while the control shrank, group sizes were never equal. Metrics were evaluated as per-user normalized rates rather than absolute counts. The permanent holdout provided a stable causal baseline regardless of the current wave size.

\textbf{Bidirectional switching as signal.} Users could move between variants freely. Declining opt-out rates over successive releases indicated improving product-market fit, independent of controlled metrics.

\textbf{Why this requires dual boot.} This phased approach was only possible because both variants coexisted in the same binary. A separate app listing would require users to actively download or delete applications, making bidirectional switching and controlled wave expansion impractical.

\begin{figure}[htbp]
    \centering
    \includegraphics[width=\columnwidth]{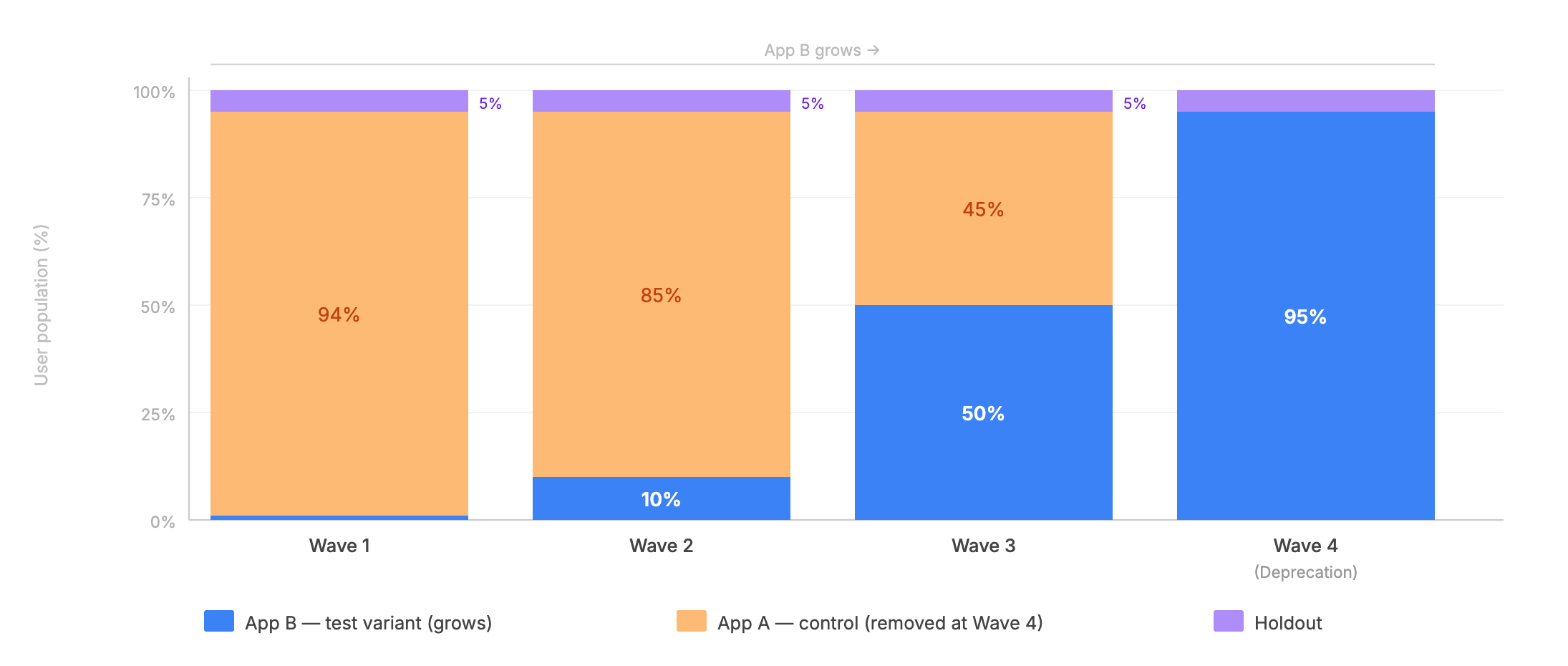}
    \caption{Phased rollout. User population shifts gradually from App A to App B across four waves.}
    \label{fig:phased-rollout}
\end{figure}

\subsection{Graceful Deprecation Without User Disruption}

Once experiments validate App B's readiness, the architecture provides a uniquely smooth deprecation path. Application state, including \texttt{NSUserDefaults}, local caches, and persisted documents, is preserved across all transitions---whether a user switches variants through opt-in and opt-out, receives a new experiment group assignment, or transitions through final deprecation.

\textbf{Properties of this deprecation model:}

\begin{itemize}
    \item \textbf{Same bundle identifier:} The app store listing never changes. Users perceive a feature update, not a product migration.
    
    \item \textbf{No user action required:} The transition is invisible---delivered as a standard app update through normal distribution channels.
    
    \item \textbf{Preserved install base:} App store ratings, reviews, download count, and search ranking are retained.
    
    \item \textbf{Minimal code change:} Deprecation requires only removing the legacy dylib from the build configuration and updating the boot logic to unconditionally load App B.
    
    \item \textbf{Reversible:} If post-deprecation metrics regress, the legacy dylib can be re-added in a subsequent release.
\end{itemize}

This stands in contrast to the conventional approach of launching a new app (new bundle ID, new listing, requiring users to discover and download a separate application), which typically results in significant user attrition during migration.

\section{Build Infrastructure and Continuous Integration}
\vspace{1em} \noindent \begin{center} \includegraphics[width=\columnwidth, height=0.38\textheight, keepaspectratio]{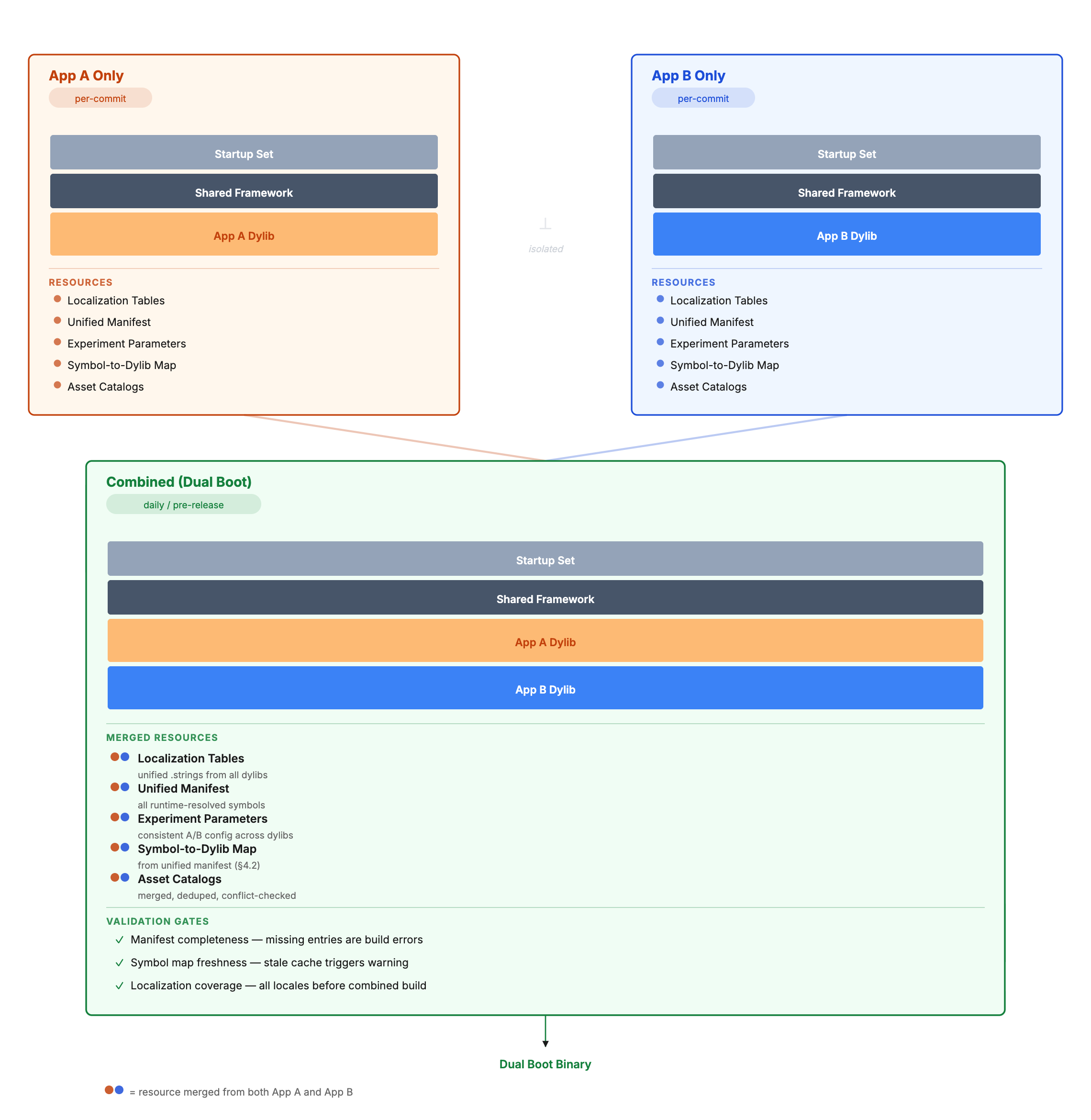} \captionof{figure}{Multi-pipeline build strategy: App A and App B build independently per-commit each with their own resources; converging into the Combined pipeline which merges all resources into a single dual-boot binary.} \label{fig:deps} \end{center} \vspace{1em} %

\subsection{Multi-Pipeline Build Strategy}

The dual boot architecture necessitates a thoughtful CI/CD strategy. A single monolithic build pipeline would negate the isolation benefits by coupling both variants' release timelines. Instead, we employ parallel build pipelines:

\begin{table}[htbp]
    \centering
    \caption{Build pipeline configuration for dual boot architecture.}
    \label{tab:build-pipelines}
    \begin{tabular}{@{}p{1.8cm}p{2.5cm}p{1.8cm}p{1.4cm}@{}}
        \toprule
        \textbf{Build Train} & \textbf{Contents} & \textbf{Purpose} & \textbf{Frequency} \\
        \midrule
        App A Only & Startup Set + Shared Framework + App A Dylib & Development iteration and release for App A & Per-commit (merge-blocking) \\
        \addlinespace
        App B Only & Startup Set + Shared Framework + App B Dylib & Rapid development iteration for App B team & Per-commit (merge-blocking) \\
        \addlinespace
        Combined (Dual Boot) & Full binary with all dylibs & Integration testing, QA sign-off, release & Daily / pre-release \\
        \bottomrule
    \end{tabular}
\end{table}

This separation allows each team to iterate at their own cadence. While monorepo strategies~\cite{ref17} centralize code, our approach centralizes only shared resources while distributing compilation. A change to App B's dylib triggers only the App B and Combined pipelines---App A's pipeline remains unaffected, preserving its build cache and validation state.

\subsection{Centrally Generated Shared Resources}

While dylibs provide code-level isolation, certain resources must be generated centrally because they are referenced across all dylibs within a single application bundle. These represent the coordination points in the architecture:

\begin{description}[leftmargin=0pt, labelindent=0pt, style=unboxed] \item[\textbf{Localization Tables.}] One localization bundle per app package; strings from all dylibs must be extracted into a single \texttt{.strings/.lproj} resource. A build script traverses all dylib targets, extracts localizable strings, deduplicates, and generates unified localization tables. \item[\textbf{Unified Manifest.}] The runtime resolution system needs a complete catalog of all runtime-resolved symbols across all dylibs. Per-dylib manifests are merged into a unified manifest during the combined build stage. \item[\textbf{Experiment Parameters.}] A/B test configurations must be consistent across dylib boundaries: a feature flag checked in the Shared Framework must resolve identically when checked in an app-specific dylib. A single configuration table is generated at build time and bundled as a shared resource accessible from any dylib. \item[\textbf{Symbol-to-Dylib Map.}] Runtime resolution requires knowledge of all symbols across all dylibs. Generated from the unified manifest (see Section~\ref{sec:build-pipeline}). \item[\textbf{Asset Catalogs.}] Images, colors, and other assets may be referenced across dylib boundaries. A merged asset catalog with deduplication and conflict detection is generated during the combined build. \end{description}

\subsection{Build Validation Gates}

To prevent release with inconsistent shared resources, the CI system enforces validation gates:

\begin{itemize}
    \item \textbf{Manifest completeness check:} Every runtime-resolved symbol referenced in code must appear in the unified manifest. Missing entries are build errors.
    
    \item \textbf{Symbol map freshness:} The combined build regenerates the symbol map from scratch; any discrepancy with a cached version triggers a warning.
    
    \item \textbf{Localization coverage:} New strings added in any dylib must have corresponding entries in all supported locales before the combined build succeeds.
    
    \item \textbf{Cross-dylib boundary validation:} Symbol interface/implementation pairs are verified to satisfy their co-location constraints (Section~\ref{sec:boundary-constraints}).
\end{itemize}

\section{Evaluation}
\label{sec:evaluation}

The architecture described in this paper was applied to an iOS application serving a large user base. The application underwent a complete ground-up rewrite using the dual boot approach: the new variant was embedded as a dynamic framework within the existing binary, validated through controlled experimentation across multiple user segments, and iterated upon over successive releases. The legacy variant was subsequently deprecated via a standard app update. Throughout the entire lifecycle, no users were lost to migration, and the app store listing, brand identity, and install base remained continuous.

The following evaluation demonstrates how the architecture addresses each challenge identified in~\S\ref{sec:introduction}:

\begin{table}[h]
\caption{How dual boot addresses identified challenges}
\label{tab:challenges-addressed}
\small
\begin{tabular}{p{2.5cm}p{3.2cm}p{2.1cm}}
\toprule
\textbf{Challenge (from \S\ref{sec:introduction})} & \textbf{How Dual Boot Addresses It} & \textbf{Evidence} \\
\midrule
Dependency graph entanglement & Each variant compiles in its own dylib with independent dependency resolution & \S\ref{sec:build-isolation}: near-zero cross-variant regressions \\
\addlinespace
Changes to one variant affect the other & Build-time isolation prevents structural cross-contamination & \S\ref{sec:build-isolation}, \S\ref{sec:deprecation-deletion}: Deprecation as Deletion \\
\addlinespace
Developers maintain mental models of multiple contexts & Each variant's codebase is invisible to the other at the source level & \S\ref{sec:quant-observations}: onboarding in days, not weeks \\
\addlinespace
Risk of a rewrite & In-binary experimentation with instant rollback & \S\ref{sec:controlled-experimentation}: controlled experimentation; \S\ref{sec:quant-observations}: zero user attrition \\
\addlinespace
User migration to a new app & Same bundle ID throughout lifecycle; deprecation via app update & \S\ref{sec:quant-observations}: brand continuity preserved \\
\bottomrule
\end{tabular}
\end{table}

\subsection{Technical Advantages}
\label{sec:technical-advantages}

\subsubsection{Build-Time Isolation}
\label{sec:build-isolation}

Both UI branching and dual boot can implement variant-level experimentation by routing users at the application entry point. The distinction is structural. With UI branching, both paths compile within the same target and share a single dependency graph. Over time, shared infrastructure accumulates implicit references to both paths, making it difficult to determine which dependencies serve the experiment and which are genuinely shared. Upon deprecation, these entangled dependencies remain as residual binary bloat.

With dual boot, each variant resolves its own dependencies within its own compilation target. Removing the experimental variant's dylib from the build removes its dependencies automatically, with no audit required. Research has shown that organizational and structural boundaries directly influence software quality~\cite{ref16}; build-time dylib isolation provides such a boundary."

A single top-level conditional that delegates to a fully rewritten code path eliminates scattered branching at the source level. However, both paths still compile within the same target: they share linker invocations, dependency resolution, and build validation. A compilation failure in the rewritten path blocks the stable path from shipping. The dependency entanglement described above applies equally, as the rewritten path's libraries are linked into the same binary regardless of whether the branch is taken at runtime.

\subsubsection{Zero Conditional Logic Pollution}
\label{sec:zero-conditional}

When a rewrite is implemented incrementally via UI branching, each modified screen or flow introduces its own conditional: the codebase accumulates branching logic across every product surface \cite{ref6} that has been partially rewritten. Each branch point must be reasoned about by every developer, covered by tests in both configurations, and maintained until deprecation. Dual boot eliminates this: the variant selection occurs once at the process entry point, and no conditional logic exists below it.

In the case of a full UI rewrite behind a single top-level branch, the source-level pollution is similarly minimal. However, the build-level coupling described in~\S\ref{sec:build-isolation} still applies: both paths share compilation, dependency resolution, and linker invocation regardless of how cleanly the branch is structured in source.

\subsubsection{Independent Technology Choices}
\label{sec:independent-tech}

With UI branching, both variants are constrained to coexist within the same compilation unit. Adopting a new UI framework, state management pattern, or dependency injection approach for the rewrite requires compatibility with the legacy code. With dual boot, each variant's dylib can adopt entirely different frameworks and patterns since they compile independently and share only the Shared Framework's stable API interface.

\subsubsection{Measurable Size Attribution}
\label{sec:size-attribution}

With UI branching, it is difficult to determine how much binary size the rewrite adds because code is interleaved in the same compilation unit. With dual boot, each dylib's size is independently measurable, enabling data-driven decisions about when carrying both variants is no longer justified.

\subsubsection{Reduced Compilation Surface}
\label{sec:reduced-compilation}

With UI branching, every developer compiles the entire application regardless of which variant they are modifying. As the rewrite grows, build times increase for all teams proportionally to the combined codebase size. With dual boot, each team builds only their variant's dylib and the shared framework during development. The combined build (assembling all dylibs into one binary) runs only during integration testing and release. Developers working on the new variant experience build times proportional to their variant's size alone, not the full application. For large-scale rewrites where the combined binary may be two to three times the size of either individual variant, this translates directly into faster iteration cycles and shorter feedback loops during development.

\subsubsection{Deprecation as Deletion}
\label{sec:deprecation-deletion}

In the dual boot architecture, deprecating the legacy variant consists of removing its dylib from the build configuration. Because the dylib constitutes a self-contained compilation unit with independently resolved dependencies, its removal eliminates all associated code and libraries from the final binary. No manual audit of shared infrastructure is required, and the resulting binary size reduction corresponds exactly to the removed dylib and its exclusive dependencies.

With UI branching, even in the single top-level branch case, removing the source code of the deprecated path does not guarantee a corresponding reduction in binary size. Dependencies introduced for the deprecated path remain in the shared dependency graph unless individually identified and excised. Shared infrastructure that was extended to support both paths requires auditing to determine which modifications remain necessary. \vspace{1em} \noindent \begin{center} \includegraphics[width=\columnwidth, height=0.375\textheight, keepaspectratio]{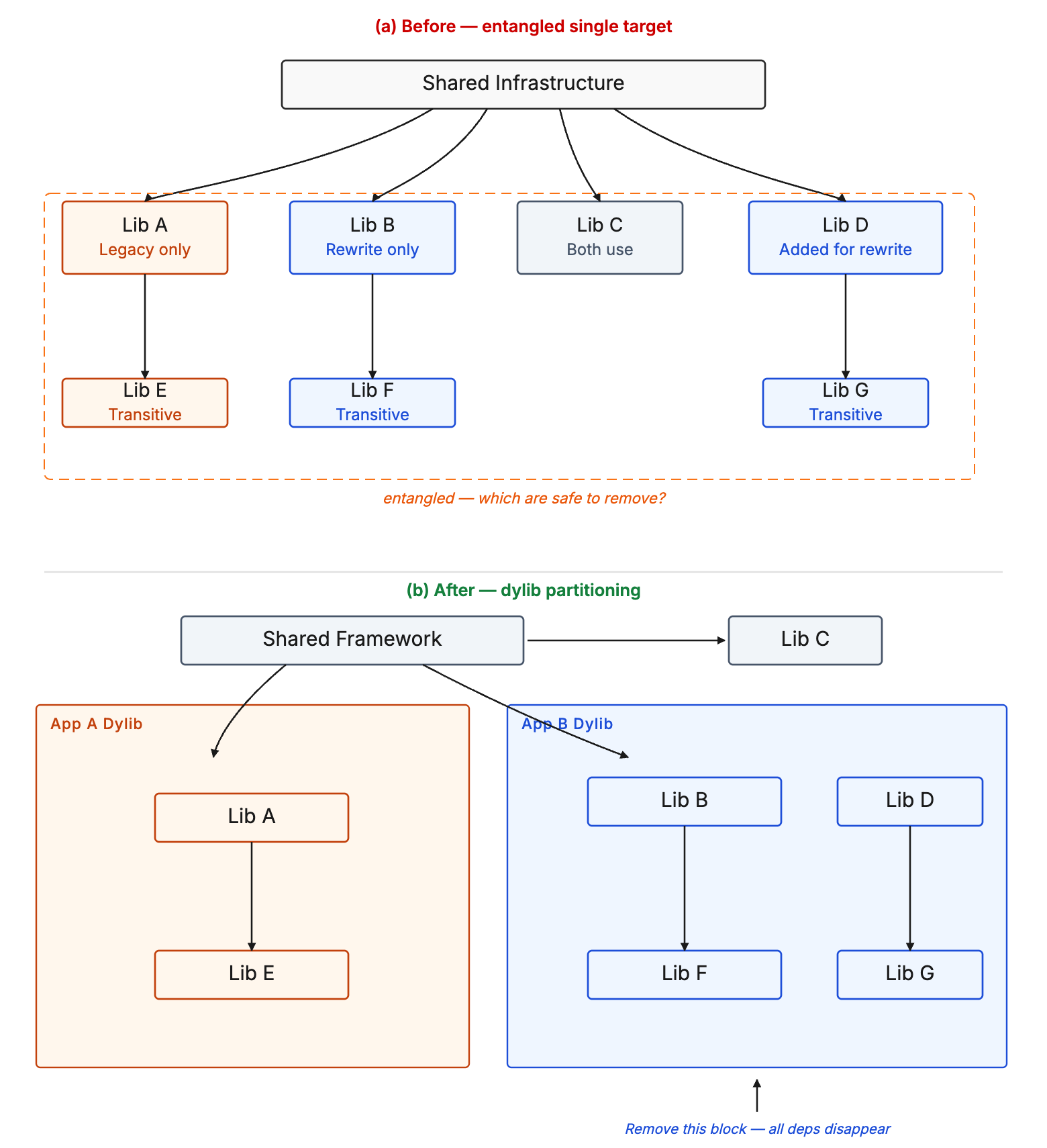} \captionof{figure}{Dependency isolation comparison. UI branching (a) creates entangled dependencies; dual boot (b) provides clean removal boundaries.} \label{fig:deps} \end{center} \vspace{1em} %

\subsection{Strategic Advantages}
\label{sec:strategic-advantages}

\subsubsection{QA Efficiency}
\label{sec:qa-efficiency}

With dual boot, each variant's dylib can be built and tested independently through its own build pipeline, enabling focused QA cycles scoped to the variant under development.

\subsubsection{Abortability Without Cleanup}
\label{sec:abortability}

The deprecation-as-deletion property (\S\ref{sec:deprecation-deletion}) applies equally when abandoning a rewrite mid-development. If the experimental variant is cancelled, removing its dylib from the build configuration eliminates all associated code with no residual impact on the primary codebase. No cleanup effort is required.

\subsubsection{Evidence-Based Migration Decisions}
\label{sec:evidence-based}

The rewrite decision does not require a commitment based on projected outcomes. Leadership observes real engagement and retention metrics from users on the new variant before authorizing deprecation. A permanent holdout group (users who never see the new variant) enables causal measurement: metric changes are attributable to the new variant specifically, isolated from external factors such as seasonality, market shifts, or operating system updates. The decision to deprecate is therefore evidence-based and causally validated.

\subsubsection{Iterative Validation Within the Existing User Base}
\label{sec:iterative-validation}

The new variant is tested on the application's actual users rather than an unrepresentative set of early adopters on a new app store listing. Product-market fit is validated iteratively, release over release, with the full behavioral diversity of the existing audience.

\subsubsection{Controlled Experimentation}
\label{sec:controlled-experimentation}

The legacy variant continues serving users not yet in the test group. Because reversion is a configuration change rather than a new app store release, the team ships with greater confidence. Features that would require extended validation in a single-shot rewrite can be released, monitored, and reverted within hours.

\subsubsection{Zero Re-Acquisition Effort}
\label{sec:zero-reacquisition}

No user acquisition effort, app store optimization effort, or ratings rebuilding is required. The transition is invisible to end users and to the app store.

\subsection{Quantitative Observations}
\label{sec:quant-observations}

\subsubsection{Business Impact}
 We list out impact of migration to business below:
 
\begin{table}[h]
\caption{Business impact }
\label{tab:business-impact}
\small
\begin{tabular}{p{2.5cm}p{5cm}}
\toprule
\textbf{Metric} & \textbf{Observation} \\
\midrule
Brand continuity & Same app store listing, bundle identifier, ratings, and reviews preserved through entire lifecycle (embed, experiment, deprecate) \\
\addlinespace
Experimentation velocity & First real-user validation achieved within days of variant code-complete, via experiment configuration in existing binary \\
\addlinespace
User attrition through deprecation & Zero. Users received the transition as a standard app update with no action required. \\
\addlinespace
Rollback capability & Instant reversion to previous variant via experiment config change; exercised multiple times during validation \\
\bottomrule
\end{tabular}
\end{table}

\newpage
\subsubsection{Developer Experience}
Impact on multi-team developer velocity:

\begin{table}[h]
\small
\begin{tabular}{p{2.5cm}p{5cm}}
\toprule
\textbf{Metric} & \textbf{Observation} \\
\midrule
Local build time (per-variant) & $\sim$0.5$\times$ compared to building the full combined binary. Each team builds only their variant during development. \\
\addlinespace
CI/CD build duration & Comparable to single-app baseline. Combined build runs only release. \\
\addlinespace
Cross-variant regressions & Near-zero, interface violations caught at build time. \\
\addlinespace
Onboarding & New engineers productive on their variant within days without needing context of the other variant \\
\addlinespace
Deprecation effort & Build configuration change only. No source code audit required. \\
\addlinespace
Shared code reuse & $\sim$40\% of total binary resided in shared dylib \\
\addlinespace
Parallel team velocity & Teams shipped independently with no merge conflicts on shared files \\
\addlinespace
Testability & Each variant's dylib tested in isolation (unit + integration) via dedicated build pipeline \\
\bottomrule
\end{tabular}
\end{table}

\subsubsection{Quality}
Impact on product quality and sentiment:

\begin{table}[h]
\small
\begin{tabular}{p{2.8cm}p{4.7cm}}
\toprule
\textbf{Metric} & \textbf{Observation} \\
\midrule
\multicolumn{2}{l}{\textit{Performance}} \\
\addlinespace
\quad Cold start (pre-main) & No measurable difference. \\
\addlinespace
\quad Cold start (post-main) & Neutral. Variant initialization adds single-digit milliseconds. \\
\addlinespace
\midrule
\multicolumn{2}{l}{\textit{Reliability}} \\
\addlinespace
\quad Crash rate & No regression attributable to dual boot infrastructure. Variant-specific crashes isolated to respective dylibs. \\
\addlinespace
\quad Stall rate & Neutral. DylibLoader resolution adds sub-millisecond latency; cached lookups are $O(1)$. \\
\addlinespace
\midrule
\multicolumn{2}{l}{\textit{App Size}} \\
\addlinespace
\quad Dual boot binary & $\sim$1.2$\times$ single-app baseline (both dylibs present, shared framework deduplicated). \\
\addlinespace
\quad Post-deprecation & Returns to $\sim$1$\times$ baseline. Legacy dylib and its exclusive dependencies removed entirely. \\
\addlinespace
\quad Symbol map size & Negligible ($<$0.1\% of binary size). JSON resource bundled in app package. \\
\addlinespace
\midrule
\multicolumn{2}{l}{\textit{Store Ratings}} \\
\addlinespace
\quad User ratings & Maintained throughout the lifecycle. Same bundle ID and listing preserved continuity; no rating reset at any phase. \\
\bottomrule
\end{tabular}
\end{table}

\subsection{Comparison of Rewrite Strategies}
\label{sec:strategy-comparison}

Table~\ref{tab:strategy-comparison} compares the three primary rewrite strategies across key metrics.

\begin{table*}[t]
\caption{Comparison of rewrite strategies}
\label{tab:strategy-comparison}
\small
\begin{tabular}{p{3.1cm}p{4cm}p{4cm}p{4cm}}
\toprule
\textbf{Metric} & \textbf{New App} & \textbf{UI Branching} & \textbf{Dual Boot} \\
\midrule
User retention through transition & \bad~Significant attrition (users must find \& download new app) & \good~No transition & \good~No transition: normal app update \\
\addlinespace
Effort to first real-user validation & \bad~High (new store listing, user acquisition, review process) & \good~Low (feature flag in existing app) & \good~Low (experiment config in existing binary) \\
\addlinespace
Rollback if rewrite fails & \bad~Catastrophic: stranded users & \yellow~Medium: tangled branch removal & \good~Trivial: config change \\
\addlinespace
Scope reduction (shared infra) & \bad~$\sim$0\%: rebuild everything & \yellow~$\sim$50--70\% reused, but polluted with branches & \yellow~$~\sim$50--70\% cleanly shared via dylib interfaces \\
\addlinespace
Developer isolation & \good~Full (separate repo) & \bad~None: shared files, merge conflicts & \good~Full: build-enforced boundaries \\
\addlinespace
Install base \& brand continuity & \bad~Start from zero & \good~Preserved & \good~Preserved \\
\bottomrule
\end{tabular}
\end{table*}

\section{Learnings}
\label{sec:learnings}

\subsection{Dependency Hygiene as an Ongoing Concern}
\label{sec:dependency-hygiene}

Maintaining strict dylib boundaries is not a one-time architectural decision but a continuous discipline. A single transitive dependency added to a shared module can pull an entire app-specific dylib into the shared framework, negating the isolation benefit. Automated dependency boundary checks at the build level are necessary but not sufficient; team education and code review norms must reinforce the boundaries.

\subsection{Dynamic Library Load Latency}
\label{sec:load-latency}

Loading a dynamic library at runtime incurs variable latency depending on system state: thread contention for runtime locks, memory pressure, and metadata registration all contribute. Observed load times ranged from single-digit milliseconds to hundreds of milliseconds. The boot-time loading strategy mitigates this by loading the variant during early initialization when contention is minimal, but post-launch lazy loading of additional libraries remains unpredictable for latency-sensitive paths.

\subsection{Symbol Namespace Management}
\label{sec:symbol-namespace}

When multiple dylibs define the same symbol, the runtime resolves it non-deterministically. This produces subtle, hard-to-reproduce bugs. The build pipeline must enforce symbol uniqueness across all dylibs and detect duplicates as build errors rather than runtime warnings.

\subsection{Specialized Build Knowledge}
\label{sec:specialized-knowledge}

Dylib-related issues (linker flags, symbol visibility, load paths, retain lists) require specialized build system expertise that most product engineers do not possess. Organizations adopting this architecture should expect dedicated build infrastructure support and developer tooling that abstracts the complexity from day-to-day development.

\subsection{Testing Infrastructure Adaptation}
\label{sec:testing-adaptation}

Standard unit and integration test infrastructure assumes a single-binary application target. Multi-dylib architectures require adaptations: per-variant test pipelines, correct install path configuration for test bundles, and separate compilation for simulator versus device architectures.

\section{Threats to Validity}
\label{sec:threats}

\paragraph{External validity}
The architecture was applied at a single organization on a single platform (iOS). While the underlying mechanisms (dynamic linking, runtime symbol resolution) are platform-standard, the build infrastructure and tooling that support the architecture are organization-specific. Replication at other organizations or on Android requires adaptation to different build systems and runtime environments. Additionally, the phased roll out strategy (§5.3) assumes reliable server-side experiment configuration delivery; under degraded network conditions, configuration fetch may introduce delays and failures \cite{ref18}, leading to incomparable experiment groups.

\paragraph{Construct validity}
Metrics such as ``cross-variant regressions: near-zero'' rely on build system enforcement rather than exhaustive runtime monitoring. It is possible that subtle cross-variant interactions exist at the shared framework boundary that are not captured by build-time validation alone.

\section{Contributions}
\label{sec:contributions}

\begin{itemize}
    \item A dylib-partitioned architecture hosting two full application variants with mutual build-time isolation and runtime code sharing via a stable shared framework interface.
    
    \item A build pipeline that generates entry-point-to-dylib resource maps and corresponding linker retain lists from a unified manifest, enabling runtime resolution of symbols that would otherwise be eliminated by dead-code optimization.
    
    \item A \texttt{DylibLoader} component that uses the generated map to resolve runtime-resolved symbols across variant dylib boundaries with lazy loading and caching.
    
    \item The first formalization of the Strangler Fig pattern for native mobile platforms, addressing single-binary distribution, app identity coupling, and linker optimization constraints absent from server-side systems.
    
    \item In-binary variant-level A/B experimentation: testing entire application architectures against each other within a single distributable binary, with symmetric exposure handling for pre-config boot decisions.
    
    \item Graceful deprecation preserving bundle identity, app store listing, ratings, and install base: achieved by removing a dylib from the build configuration rather than migrating users to a new application.
\end{itemize}

\section{Conclusion}
\label{sec:conclusion}

The dual boot architecture demonstrates that a mobile application rewrite need not be a high-stakes, irreversible commitment. By combining dynamic library partitioning with build-generated runtime-resolved symbol resolution, organizations gain the ability to embed, experiment, iterate, and deprecate within a single binary, transforming what has historically been an all-or-nothing bet into a controlled, measurable, reversible process.

We have presented a complete architecture for operating multiple mobile application experiences from a shared codebase, framed as the Strangler Fig pattern adapted for native mobile constraints. The system covers the full rewrite lifecycle: experimentation (in-binary A/B testing), validation (opt-in/opt-out with rollback), and deprecation (dylib removal preserving app identity). Unlike industrial precedents---Uber's big-bang cutover, Google's same-app modularization, Airbnb's flag-based migration---dual boot uniquely combines developer isolation with user continuity, clean abortability, and zero conditional logic. The approach generalizes to any platform with dynamic library support and any runtime extension mechanism.

\bibliographystyle{ACM-Reference-Format}
\bibliography{references}

\end{document}